\documentclass[aps,prb,twocolumn,preprintnumbers,showpacs,superscriptaddress]{revtex4}

\usepackage{graphicx}
\usepackage{tabularx}
\usepackage{amssymb}
\usepackage{amsmath}
\usepackage{bm}
\usepackage{subfigure}
\usepackage{graphpap}
\usepackage{multirow}
\usepackage{notes2bib}
\usepackage[version=3]{mhchem} 

\begin{document}
\title{Optical spectra from molecules to crystals: Insight from many-body perturbation theory}
\author{Caterina \surname{Cocchi}}
\affiliation{Institut f\"ur Physik and IRIS Adlershof, Humboldt-Universit\"at zu Berlin, Berlin, Germany}
\affiliation{European Theoretical Spectroscopic Facility (ETSF)}
\email{caterina.cocchi@physik.hu-berlin.de}
\author{Claudia \surname{Draxl}}
\affiliation{Institut f\"ur Physik and IRIS Adlershof, Humboldt-Universit\"at zu Berlin, Berlin, Germany}
\affiliation{European Theoretical Spectroscopic Facility (ETSF)}
\date{\today}
\pacs{71.15.Mb, 71.35.Cc, 78.40.Me}
\begin{abstract}
Time-dependent density-functional theory (TDDFT) often successfully reproduces excitation energies of finite systems, already in the adiabatic local-density approximation (ALDA). Here we show for prototypical molecular materials, i.e., oligothiophenes, that ALDA largely fails and explain why this is so. By comparing TDDFT with an in-depth analysis based on many-body perturbation theory, we demonstrate that correlation effects crucially impact energies and character of the optical excitations not only for molecules of increasing length and in crystalline environment, but even for isolated small molecules. We argue that only high-level methodologies, which explicitly include correlation effects, can reproduce optical spectra of molecular materials with equal accuracy from gas phase to crystal structures.
\end{abstract}
\maketitle
\section{Introduction}
Optical excitations in organic materials are strongly dominated by many-body effects. Electron-electron interactions determine the electronic structure, and electron-hole (\textit{e-h}) correlations rule the excitation process.
A methodology that is able to consistently capture the features of molecular materials, from single molecules to their condensed phases, including organic crystals,\cite{schw+07book,buss+02apl,pusc-ambr02prl,ruin+02prl,humm+04prl,humm-ambr05prb,sai+08prb} adsorbate systems, \cite{taut07pss,wegn+08nl} hybrid materials and nanostructures, \cite{blum+06prl,zahn+07cr} is an essential prerequisite to predict their excited-state properties.
Many-body perturbation theory (MBPT) represents the state-of-the-art method to calculate optical excitations in solids. \cite{onid+02rmp}
The \textit{GW} approach \cite{hedi65pr} gives quasi-particle (QP) energies and the solution of the Bethe-Sapleter equation (BSE) \cite{hank-sham80prb,stri88rnc} yields excitation energies and wavefunctions of the \textit{e-h} pairs.
Although, in principle, $GW$+BSE can be applied to any material, it is computationally too demanding for many systems of technological interest.
Quantum chemistry offers powerful tools, such as Coupled Clusters (CC) and Configuration Interaction Singles and Doubles (CISD), \cite{jens07book} to accurately compute optical properties of molecules.
Again, only small systems can be treated with these methods, due to their huge computational costs. 
 
Since the turn of the century, time-dependent density-functional theory (TDDFT) has gained increasing popularity, \cite{burk+05jcp} due to its remarkable ability in reproducing optical spectra of small molecules and clusters, with relatively low computational effort. \cite{yaba-bert99pra1,vasi+02prb,furc-ahlr02jcp,tiag-chel05ssc,tiag-chel06prb}
This success is mainly ascribed to the dominance of the long-range part of the Coulomb potential $v$ over the exchange-correlation kernel ($f_{xc}$). \cite{sott+05ijqc}
$v$ blue-shifts the absorption onset above the Kohn-Sham (KS) gap, and redistributes the oscillator strength (OS) to higher energies, providing a good approximation for the optical spectra. \cite{onid+02rmp}
However, TDDFT suffers from severe drawbacks when dealing with extended systems.
The spurious long-range behavior of standard $f_{xc}$ inhibits reproducing bound excitons in solids. \cite{ghos+97prb,bott+07rpp}
For the same reason, TDDFT is unable to describe charge-transfer-like excitations in molecular complexes. \cite{grim-para03chpch,dreu+03jcp,toze03jcp,mait05jcp,auts09chpch}
To overcome these limitations, new kernels have been developed including many-body effects \cite{rein+02prl,dels+03prb,mari+03prl1,bott+04prb,turk+09prb} and exhibiting the correct long-range behavior. \cite{vanf+02prl,grit-baer04jcp,kuri+11jctc}
Unfortunately, these improvements mainly concern specific classes of materials and/or excitations, and therefore do not often extend, in practice, the range of applicability of TDDFT.

The goal of this work is to understand the role of many-body effects in the optical excitations of molecular materials, from the gas-phase to crystals, and to clarify when and why TDDFT can be trusted.
We adopt the adiabatic local-density approximation (ALDA), \cite{zang-sove80pra} as the simplest and most common kernel of TDDFT.
While more sophisticated kernels can quantitatively improve the results, \cite{bott+07rpp,peac+08jcp,jacq+09jctc,laur-jacq13ijqc,sun-auts14jctc} the physical picture is already clear from ALDA. \cite{dreu-head05cr}
We compare TDDFT with MBPT, with the aim to identify the most relevant contributions to the optical excitations and to understand to which extent TDDFT is able to reproduce them.
To do so, we choose the family of oligothiophenes as prototypical example.
Since only a few intense peaks characterize their UV-visible spectra, \cite{fich08book} they are ideally suited for this study.
We investigate thiophene oligomers with an even number of rings, up to 6, as well as the single thiophene ring, going from the isolated molecule to the crystal.

\section{Theoretical Background and Computational Details}
Optical absorption spectra presented in this work are computed from frequency-dependent linear-response TDDFT, as well as from MBPT, in the framework of $G_0W_0$+BSE. \cite{rohl-loui00prb}$^,$ \bibnote{The QP correction to the gap, computed from first-principles within the $G_0W_0$ approximation, is applied to the LDA electronic structure through a scissor operator.}

In TDDFT, optical excitations are calculated from the solution of the Dyson-like equation for the dynamic polarizability $\chi = \chi_0 + \chi_0 (v+f_{xc}) \chi$,
where $\chi_0$ is the KS response function.
In ALDA, the kernel $f_{xc}^{ALDA}$ is local and static. \cite{zang-sove80pra,ghos+97prb}
For $f_{xc}=0$, $\chi$ is computed within the random-phase approximation (RPA), while neglecting also $v$ leads to the independent-particle approximation (IPA), with $\chi=\chi_0$.
The dielectric tensor $\epsilon$ is calculated as $\epsilon_{\mathbf{G,G'}}(\mathbf{q},\omega) = \delta_{\mathbf{G,G'}} - v_{\mathbf{G}}(\mathbf{q}) \chi_{\mathbf{G,G'}}(\mathbf{q},\omega)$.

Optical excitations are computed from MBPT through a two-step procedure. \cite{rohl-loui00prb}
First, the QP correction is obtained from $G_0W_0$.
Then, exciton energies are given by the solution of the BSE, in the matrix form: $H^{BSE} A^{\lambda} = E^{\lambda} A^{\lambda}$.
The effective two-particle Hamiltonian $H^{BSE} = H^{diag} + 2 \gamma_x H^x + \gamma_c H^{dir}$ is composed of three terms, which can be \textit{switched on} and \textit{off}, depending on the values of the coefficients $\gamma_x$ and $\gamma_c$. 
For $\gamma_x$=$\gamma_c$=1 ($\gamma_x$=0, $\gamma_c$=1) \textit{singlet} (\textit{triplet}) excitations are calculated.
The \textit{diagonal} term $H^{diag}$ accounts for single-particle transitions.
The repulsive $e$-$h$ \textit{exchange} term $H^x$, including the short-range Coulomb interaction, describes local-field effects (LFE).
The \textit{direct} term $H^{dir}$ contains the screened Coulomb interaction, which determines the attractive $e$-$h$ interaction.
The imaginary part of the macroscopic dielectric function is related to the BSE eigenvectors $A^{\lambda}$, which correspond to the amplitude of the excitons.
They also carry information about the character of the excitons and the weight of the single-particle transitions contributing to them.

All calculations are performed with the \texttt{exciting} code, \cite{gula+14jpcm} a computer package for density-functional theory and MBPT, implementing the all-electron full-potential augmented planewave method.
The KS electronic structure is computed using the Perdew-Wang local-density approximation (LDA) as the exchange-correlation ($xc$) functional. \cite{perd-wang92prb}
In \texttt{exciting}, TDDFT and BSE are treated on the same footing, \cite{sagm-ambr09pccp} enabling a direct comparison between the results.
Isolated molecules are considered in orthorhombic supercells, including at least 7 \AA{} of vacuum in each lattice direction.
The internal coordinates are relaxed, with a threshold of 0.025 eV/\AA{} for the interatomic forces.
All the resulting geometries are flat.
For the calculation of $\chi$, 100 empty states are included.
In $G_0W_0$ calculations, the dynamically screened Coulomb potential $W_0$ is computed within RPA, including 100 empty states for the crystalline structures and 500 for the isolated molecules.
For the solution of the BSE, 500 empty states are considered to compute the screened Coulomb interaction.
The adopted computational parameters ensure accuracy of 0.05 eV for the lowest-energy excitations in the spectra.

\section{Results and Discussion}
\subsection{Molecules of increasing length}
%
\begin{figure}
\centering
\includegraphics[width=.45\textwidth]{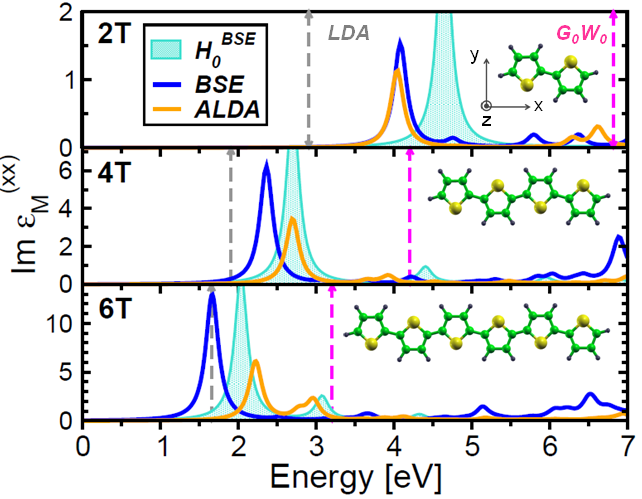}%
\caption{(Color online) Imaginary part of the macroscopic dielectric function of 2T, 4T, and 6T for light polarization parallel to the long molecular axis ($x$). Results from ALDA, BSE, and an approximation of the \textit{e-h} interaction term ($H_0^{BSE}$, see text) are presented. LDA and $G_0W_0$ gaps are indicated by dashed lines. The molecules are shown as insets with C (S) atoms depicted in green (yellow) and H in black.  A Lorentzian broadening of 0.1 eV is applied to all the spectra.  
}
\label{fig1}
\end{figure}

In Fig. \ref{fig1} we show the optical spectra of bithiophene (2T), quarterthiophene (4T) and sexithiophene (6T).
For the smallest oligomer, 2T, the ALDA spectrum is in excellent agreement with the $G_0W_0$+BSE result. 
The UV-visible region is dominated by a strong peak, polarized along the long ($x$) axis of the molecule, \cite{fabi+05jpca} and governed by the $\pi$-$\pi^*$ transition between the highest-occupied molecular orbital, HOMO (H), and the lowest-unoccupied molecular orbital, LUMO (L).
Our finding, indicating the lowest excitation energy of 2T at 4.05 -- 4.09 eV (see Table \ref{table1}), is in agreement with experiments, \cite{jone+90jpch,lap+97jpca,hick+00jacs,seix+00jpca,hutc+02jpca} and with quantum-chemistry results, \cite{rubi+03chpch,andr-wite11tcacc} which evaluate the peak between 4.05 and 4.11 eV.
As the size of the oligomer increases, discrepancies between the $G_0W_0$+BSE and ALDA spectra emerge in both the energy of the first peak and the overall spectral shape.

\begin{table*}
\begin{tabular}{c|c c c|c c c|c|c|c}
 & $E_{gap}^{LDA}$ & $E_{gap}^{G_0W_0}$ & $\Sigma-V_{xc}$ & $E_{H_0^{BSE}}$ & $E_{triplet}$ & $E_{singlet}$ & $\bm{\Delta^{MBPT}}$ & $E^{ALDA}$ & $\bm{\Delta^{TDDFT}}$ \\ 
\hline  \hline
\multirow{2}{*}{\textbf{2T}} & \multirow{2}{*}{2.91} & \multirow{2}{*}{6.82} & \multirow{2}{*}{3.91} & 4.65 & 2.19 & 4.09 & \multirow{2}{*}{\textbf{1.18}} & \multirow{2}{*}{4.05} & \multirow{2}{*}{\textbf{1.14}} \\ 
 & & & & (\textit{-2.17}) & (\textit{-4.63}) & (\textit{-2.73}) & & & \\ \hline
\multirow{2}{*}{\textbf{4T}} & \multirow{2}{*}{1.91} & \multirow{2}{*}{4.21} & \multirow{2}{*}{2.30} & 2.70 & 0.73 & 2.37 & \multirow{2}{*}{\textbf{0.46}} & \multirow{2}{*}{2.70} & \multirow{2}{*}{\textbf{0.79}} \\ 
 & & & & (\textit{-1.51}) & (\textit{-3.48}) & (\textit{-1.84}) & & & \\ \hline
\multirow{2}{*}{\textbf{6T}} & \multirow{2}{*}{1.66} & \multirow{2}{*}{3.21} & \multirow{2}{*}{1.54} & 2.04 & 0.28 & 1.66 & \multirow{2}{*}{\textbf{0.01}} & \multirow{2}{*}{2.22} & \multirow{2}{*}{\textbf{0.56}} \\ 
 & & & & (\textit{-1.17}) & (\textit{-2.92}) & (\textit{-1.54}) & & & \\
 \hline \hline
 \multirow{2}{*}{\textbf{1T crystal}} & \multirow{2}{*}{4.42} & \multirow{2}{*}{8.34} & \multirow{2}{*}{3.92} & 7.17 & 3.75 & 5.80 & \multirow{2}{*}{\textbf{1.38}} & \multirow{2}{*}{5.44} & \multirow{2}{*}{\textbf{1.02}} \\ 
 & & & & (\textit{-1.17}) & (\textit{-4.59}) & (\textit{-2.54}) & & & \\ \hline
\multirow{2}{*}{\textbf{1T expanded crystal}} & \multirow{2}{*}{4.63} & \multirow{2}{*}{9.47} & \multirow{2}{*}{4.84} & 7.69 & 3.81 & 5.77 & \multirow{2}{*}{\textbf{1.14}} & \multirow{2}{*}{5.63} & \multirow{2}{*}{\textbf{1.00}} \\ 
 & & & & (\textit{-1.78}) & (\textit{-5.66}) & (\textit{-3.70}) & & & \\ \hline
\multirow{2}{*}{\textbf{1T molecule}} & \multirow{2}{*}{4.59} & \multirow{2}{*}{9.85} & \multirow{2}{*}{5.26} & 7.14 & 3.78 & 5.85 & \multirow{2}{*}{\textbf{1.26}} & \multirow{2}{*}{6.66} & \multirow{2}{*}{\textbf{2.08}} \\ 
 & & & & (\textit{-2.71}) & (\textit{-6.07}) & (\textit{-4.00}) & & & \\
\hline \hline
\end{tabular} 
\caption{Energies relevant for the analysis of excitations in all systems considered in this work: Fundamental gaps, obtained by LDA and $G_0W_0$ and their difference ($\Sigma-V_{xc}$); exciton energies of the first \textit{triplet} ($E_{triplet}$) and \textit{singlet} ($E_{singlet}$) peak as well as from an approximation to the \textit{e-h} interaction ($E_{H_0^{BSE}}$, see text); $\Delta^{MBPT}=(\Sigma-V_{xc})+E_b$ for the first bright exciton from BSE; $\Delta^{TDDFT} = E^{ALDA}-E_{gap}^{LDA}$, with $E^{ALDA}$ being the first peak energy from TDDFT. Exciton binding energies ($E_b$) are given in parenthesis. All energies are expressed in eV.}
\label{table1}
\end{table*}

To understand the source of this disagreement, we inspect in detail the results from MBPT concerning the first excitonic peak.
This analysis is summarized in Table \ref{table1}.
Two types of many-body effects come into play: the QP correction to the electronic structure ($\Sigma-V_{xc}$) and the \textit{e-h} interaction.
Like the fundamental gap computed from LDA ($E_{gap}^{LDA}$) and $G_0W_0$ ($E_{gap}^{G_0W_0}$), also $\Sigma-V_{xc}$, which corresponds to their difference, decreases with increasing oligomer length.
Its effect is to blue-shift the absorption onset in absence of \textit{e-h} correlation and LFE.
From the solution of the BSE, the exciton binding energies ($E_b$) are computed; for simplicity, we define $E_b$ as the difference between the excitation energy and $E_{gap}^{G_0W_0}$. \cite{rohl-loui00prb,comment}
Due to its bound and localized character, the lowest exciton in 2T has a large binding energy of almost 3 eV (Table \ref{table1}).
The peak position arises from a \textit{partial} cancellation between $\Sigma-V_{xc}$ and $E_b$.
We label this difference $\Delta^{MBPT}=(\Sigma-V_{xc}$) + $E_b$, which is usually a positive quantity.
This compensation should be mimicked by the ALDA kernel, in order to correctly yield the absorption features.
For this purpose, we define $\Delta^{TDDFT}$=$E^{ALDA}$ - $E_{gap}^{LDA}$, where $E^{ALDA}$ denotes the lowest-energy bright excitation from ALDA.
The closer $\Delta^{TDDFT}$ is to $\Delta^{MBPT}$, the better the 
partial compensation between $\Sigma-V_{xc}$ and $E_b$ is reproduced by ALDA.
In 2T, $\Delta^{TDDFT} \simeq \Delta^{MBPT}$: ALDA mimics almost perfectly the blue-shift caused by the QP correction and the red-shift due to exciton binding.
As the size of the oligomer increases, the spread between $\Delta^{TDDFT}$ and $\Delta^{MBPT}$ becomes larger. 

We note that we have to deal with two different issues here. 
One is a shortcoming of ALDA, the main focus of this work, which we discuss below. 
At this point, however, we also need to critically assess the MBPT results in view of the $GW$ starting-point problem.
$G_0W_0$+BSE is known to underestimate excitation energies of gas-phase molecules, \cite{jacq+15jctc,hiro+15prb,brun+15condmat} and this shortcoming becomes more serious for large compounds.
Indeed, we find the absorption onsets of 4T and 6T to be red-shifted compared to the experimental ones, assigned at 3.2 eV and 2.9 eV, respectively. \cite{lap+97jpca,hutc+02jpca}
This issue is predominantly ascribed to the $G_0W_0$ step, \cite{jacq+15jctc,brun+15condmat} whose results may crucially depend on the underlying $xc$ functional, \cite{blas+11prb,koer-maro12prb,hues+13prb} and becomes very critical in case of large oligomers.
In Ref. \onlinecite{blas+11prb}, $G_0W_0$ gaps, computed on top of different $xc$ functionals, are presented for a number of compounds, including oligoacenes. 
Overall, LDA as starting point leads to a considerable underestimation of the $G_0W_0$ gaps compared to experiments, with the smallest oligomer described best.
A systematic analysis on the starting-point dependence of MBPT spectra in molecular materials from gas to crystal phase is definitely due, but goes beyond the scope of this work.
This issue, although affecting the agreement with experiments, does not alter exciton binding energies and excitation character.
As such, it does not impact the analysis that follows.

\begin{table}
\begin{tabular}{c|c|c|c}
 & \textbf{2T} & \textbf{4T} & \textbf{6T} \\
 \hline \hline
 $H_0^{BSE}$ & H $\rightarrow$ L (100$\%$) & H $\rightarrow$ L (100$\%$) & H $\rightarrow$ L (100$\%$) \\ \hline 
 \multirow{2}{*}{\textit{triplet}} &  \multirow{2}{*}{H $\rightarrow$ L (96$\%$)} &  \multirow{2}{*}{H $\rightarrow$ L (89$\%$)} & H $\rightarrow$ L (59$\%$) \\
 & & & H-1 $\rightarrow$ L+1 (27$\%$) \\ \hline
 \multirow{2}{*}{\textit{singlet}} &  \multirow{2}{*}{H $\rightarrow$ L (84$\%$)} &  \multirow{2}{*}{H $\rightarrow$ L (89$\%$)} & H $\rightarrow$ L (74$\%$) \\
 & & & H-1 $\rightarrow$ L+1 (20$\%$) \\
 \end{tabular} 
\caption{Single-particle transitions contributing to the first peak of 2T, 4T and 6T. $H_0^{BSE}$, \textit{triplet} and \textit{singlet} excitation energies are shown.}
\label{table2}
\end{table}

We focus on the BSE results, which help us to clarify the shortcomings of ALDA.
To this extent, we analyze the nature of the low-energy excitations in terms of single-particle contributions, namely, whether they stem from one transition or from a combination of two or more.
For this purpose, we indicate in Table \ref{table2} the composition of the lowest-energy \textit{singlet} and \textit{triplet} excitation of 2T, 4T and 6T.
While for the smaller molecules the first exciton stems from the H $\rightarrow$ L transition, regardless of whether we include the exchange and direct $e$-$h$ interaction, for 6T the lowest-energy excitation, still presenting $\pi$-$\pi^*$ character, is composed by a mixing of H $\rightarrow$ L (74$\%$) and H-1 $\rightarrow$ L+1 (20$\%$) transitions.
The same result is obtained also when the BSE exchange term ($H^x$), and hence LFE, are neglected (\textit{triplet}).
From this we can assert that \textit{e-h} correlation effects, given by $H^{dir}$, crucially affect the composition of the first exciton in 6T.
Fig. \ref{fig1} also shows the spectrum obtained from an effective approximation to the \textit{e-h} interaction term ($H_0^{BSE}$, shaded area).
This corresponds to considering only the $\mathbf{G}=\mathbf{G'}=0$ term of $H^{dir}$ (including a $\mathbf{q}$-independent screened \textit{e-h} interaction).
The Hamiltonian then becomes $H_0^{BSE} = H^{diag} + h_0$, where $h_0$ is the $\mathbf{G}=0$ term of $H^{dir}$:
\begin{equation}
h_0^{v c \mathbf{k}, v' c' \mathbf{k'}} = - \dfrac{2}{(2\pi)^2} \left(\dfrac{6\pi^2}{\Omega}\right)^{\frac{1}{3}} \dfrac{4\pi}{\epsilon(\mathbf{q = 0})} \delta_{c c'} \delta_{v v'} .
\label{eq:h0}
\end{equation}
For details on the derivation of Eq. \ref{eq:h0}, see Ref. \onlinecite{pusc-ambr02prb}.
The exchange term $H^x$ vanishes, as it contains only the short-range part of the bare Coulomb potential $\bar{v}=v-v_0$, $v_0$ being the $\mathbf{G}=0$ component of $v$.
Excitations resulting from $H_0^{BSE}$ consist of pure single-particle transitions due to $H^{diag}$, which are rigidly red-shifted by the ``zero-order" \textit{e-h} interaction from $h_0$ (see Eq. \ref{eq:h0}). \cite{pusc-ambr02prb}
In the $H_0^{BSE}$ spectrum of 6T, in addition to the intense peak at 2 eV, given by H $\rightarrow$ L, a weak peak at about 3 eV stems from the H-1 $\rightarrow$ L+1 transition (Table \ref{table2}). 
The same feature appears also in the ALDA spectrum (orange line).
This analogy, supported by the insight from the BSE results (dominance of $H^{dir}$), confirms that the local and frequency-independent $f_{xc}^{ALDA}$ cannot properly account for correlation effects.
In smaller molecules like 2T and 4T, the OS in the corresponding energy range is concentrated only in the first peak, which is indeed largely dominated by H $\rightarrow$ L (Table \ref{table2}). 
We conclude that the agreement between BSE and ALDA in the small oligomers is determined by the single-particle character of the first intense peak.
As the size of the molecule increases, correlation effects become more relevant: TDDFT can only reproduce the energy of the first excitation but not the overall spectral shape.
The description of these excitations can be improved to some extent
by adopting more sophisticated $xc$ functional and kernels, such as global or range-separated hybrids.
In fact, they improve the underlying density/wave-functions through enhancing the exchange term and thus reducing the self-interaction error.
This argument is supported by the findings of Ref. \onlinecite{sun-auts14jctc}, where the nature of the lowest-energy excitation of 6T is analyzed.
Within B3LYP, the excitation corresponds to an almost pure H $\rightarrow$ L transition.
Using long-ranged corrected PBE0 (LC-PBE0), its mixed character emerges, with the weight of H-1 $\rightarrow$ L+1 being 7$\%$.
This kernel leads to a better description of the excitation, by reproducing its mixed character. 

\begin{figure}
\centering
\includegraphics[width=.48\textwidth]{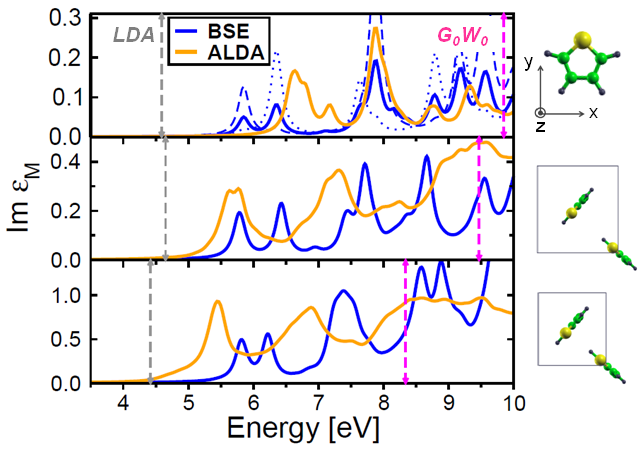}%
\caption{(Color online) BSE and ALDA spectra of the 1T molecule (top), expanded crystal (middle), and crystal (bottom), as shown on the right. Solid lines represent the average over the three Cartesian components. In the top panel, dotted (dashed) curves indicate the $xx$ ($yy$) component of Im $\varepsilon_M$ from BSE. LDA and $G_0W_0$ gaps are indicated by dashed lines. A Lorentzian broadening of 0.1 eV is applied to all the spectra.
}
\label{fig2}
\end{figure}

We have shown so far that TDDFT can yield, in excellent agreement with MBPT, the absorption spectra of small molecules like 2T, which exhibit a first intense peak with mainly single-particle character.
In this case, ALDA correctly describes the peak and mimics the partial cancellation between $\Sigma - V^{xc}$ and $E_b$, providing accurate excitation energies.
On the contrary, when correlation effects play an important role, as in 6T, TDDFT suffers from apparent shortcomings.
This is, however, not the only limiting scenario.  
The single thiophene ring (1T), despite its small size, is another problematic case for TDDFT.
In Fig. \ref{fig2} (upper panel), we show the spectrum of this molecule computed from ALDA and $G_0W_0$+BSE. 
Let us focus on the low-energy part first. 
It is characterized by two peaks, polarized along the $y$ and $x$ axis of the molecule, respectively. 
Contrary to the case of $n$T ($n$=2, 4, 6), where the H $\rightarrow$ L transition yields the lowest-energy peak, the first exciton of 1T stems from a combination of different single-particle contributions, including H-1 $\rightarrow$ L (60$\%$) and H $\rightarrow$ L+3 (30$\%$).
The second exciton is $x$-polarized and mostly due to H $\rightarrow$ L (84$\%$). 
This information is summarized in Table \ref{table3}.
Our BSE results are in good agreement with experiments and CC calculations,  \cite{holl+14pccp} while the ALDA spectrum is poor.
The first two excitations around 6.5 eV are roughly degenerate in energy.
Through indirect information about the single-particle transitions contributing to the ALDA excitations, we can attribute this inaccuracy to a lack of correlation in TDDFT, in analogy with the previous analysis of 6T.
This is also confirmed in a recent quantum-chemical study, showing that only those methods, which accurately treat correlation effects, such as CC2, EOM-CCSD, and CIS(D), correctly yield energy and character of the first two excitations (5.7 and 6.2 -- 6.4 eV, respectively). \cite{prlj+14jpcl}

\begin{table}[b]
\begin{tabular}{c|c|c|c|c} \hline \hline
 & \multicolumn{2}{c|}{$1^{st}$ excitation} & \multicolumn{2}{c}{$2^{nd}$ excitation} \\ \hline 
  &  pol. & composition & pol. & composition \\ \hline \hline
 $H_0^{BSE}$ & $x$ & H $\rightarrow$ L (100$\%$) & $y$ & H-1 $\rightarrow$ L (100$\%$) \\  \hline
 \textit{triplet} & $x$ & H $\rightarrow$ L (99$\%$) & $y$ & H-1 $\rightarrow$ L (98$\%$) \\ \hline
  \multirow{2}{*}{\textit{singlet}} & \multirow{2}{*}{$y$} & H-1 $\rightarrow$ L (60$\%$) & \multirow{2}{*}{$x$} & \multirow{2}{*}{H $\rightarrow$ L (84$\%$)} \\
  & &  H $\rightarrow$ L+3 (30$\%$) & & \\ \hline \hline
 \end{tabular} 
\caption{Analysis of the first and second excitation of the 1T molecule. Polarization direction (pol.) and composition in terms of single-particle transitions are provided for $H_0^{BSE}$, \textit{triplet}, and \textit{singlet} results.}
\label{table3}
\end{table}

\begin{figure}
\centering
\includegraphics[width=.47\textwidth]{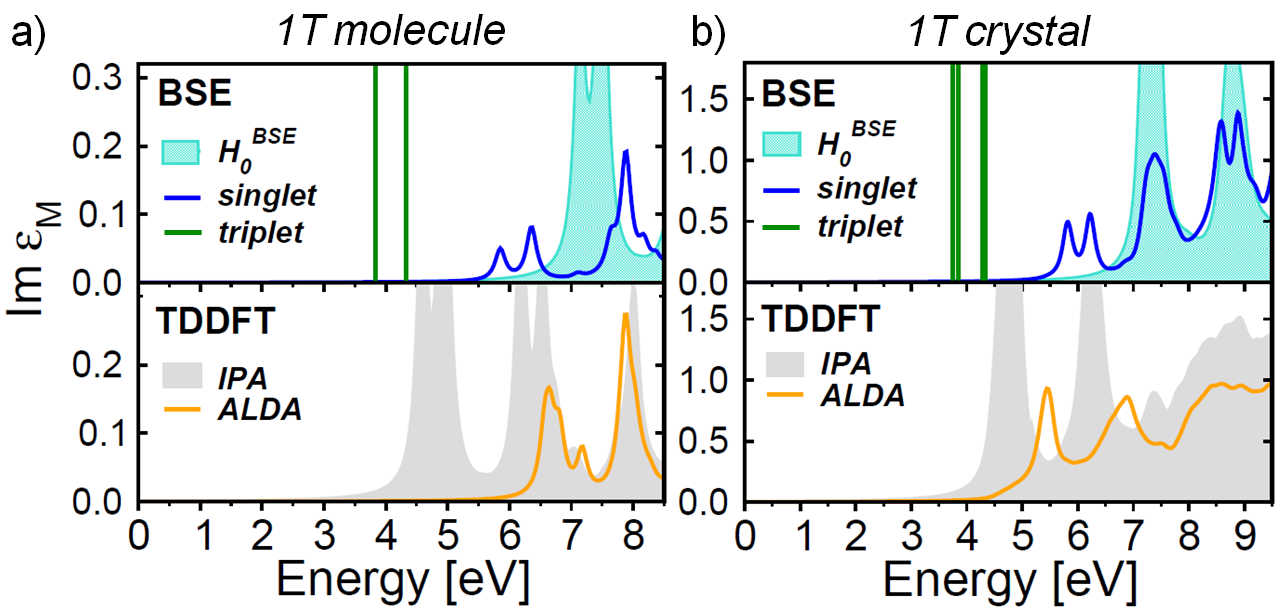}%
\caption{(Color online) Imaginary part of the macroscopic dielectric function, averaged over the three Cartesian components of 1T a) molecule, and b) crystal, computed from BSE (top) and TDDFT (bottom). Energies of the lowest-lying triplet excitations are indicated by vertical green lines. A Lorentzian broadening of 0.1 eV is applied to all the spectra.
}
\label{fig3}
\end{figure}

To better identify the cause of the disagreement between ALDA and MBPT results, we further analyze the spectra in Fig. \ref{fig3}a (upper panel). 
Compared to the $H_0^{BSE}$ spectrum (turquoise shaded area), the onset of \textit{triplet} excitation energies (green bars) is red-shifted.
We recall that in both cases the BSE Hamiltonian does not include the exchange term $H^x$.
Hence, $H_0^{BSE}$ and \textit{triplet} differ only by the treatment of the screened \textit{e-h} interaction: in the former case, it is $\mathbf{q}$-independent (see Eq. \ref{eq:h0}), while in the latter one the full dielectric tensor is included. \cite{pusc-ambr02prb} 
Remarkably, the character of lowest-energy excitations is the same (Table \ref{table3}).
From this we conclude that, no matter how we treat the attractive \textit{e-h} term, the lowest-lying excitons keep their character. 
Transferring this knowledge to the \textit{singlet} spectrum (blue line), we emphasize that only the inclusion of the exchange term, $H^x$, yields the correct description of the first two excitations.
This is in contrast to the previous example of 6T, where \textit{e-h} correlation plays the dominant role. 
In 1T, LFE are not only responsible for the redistribution of the OS to higher energies, but also for the right order and composition of the lowest excitons.
Again, ALDA cannot account for these many-body effects, inaccurately describing the first two peaks and slightly overestimating the absorption onset (see $\Delta^{MBPT}$ and $\Delta^{TDDFT}$ in Table \ref{table1}).
As shown in Fig. \ref{fig3}a (bottom panel), the main effect of the ALDA kernel (orange line), compared to the independent-particle approximation (IPA, shaded gray area), is to blue-shift the energy of the first two peaks and to redistribute the OS to higher energies.
Overall, the TDDFT results reflect the discrete spectrum of molecular levels.
Therefore, above the onset (7 -- 10 eV), ALDA captures transitions between such localized single-particle states, in better agreement with BSE than in the regime dominated by strongly bound excitons.

%

\subsection{From molecule to solid}

When intermolecular interactions come into play, the shortcomings of TDDFT become even more dramatic.
To extend our analysis in this direction, we go systematically to the 1T crystalline phase,
considered with two inequivalent molecules in an orthorhombic unit cell, of lattice vectors $a_0$=9.76 \AA{}, $b_0$=7.2 \AA{}, and $c_0$=6.67 \AA{}. \cite{abra52acry}
We investigate also a model system with lattice vectors scaled by a factor 1.25 with respect to $a_0$, $b_0$ and $c_0$: we refer to this structure as \textit{expanded} crystal.
\bibnote{A $4\times4\times4$ ($2\times2\times2$) \textbf{k}-point mesh is adopted in all the calculations of the packed (expanded) crystal.}
The corresponding spectra are shown in Fig. \ref{fig2} (bottom and middle panels).
Due to the small size of the molecular constituents, large values of $E_{gap}^{G_0W_0}$, and hence of $\Sigma-V_{xc}$ and of $E_b$, are obtained from MBPT (Table \ref{table1}).
A systematic decrease of these quantities, going from the isolated compound to its crystalline phases, indicates enhanced screening together with delocalization of the exciton wave-function, as also found for other molecular crystals and polymers. \cite{buss+02apl,pusc-ambr02prl,ruin+02prl,humm+04prl,humm-ambr05prb}
A number of intense bound excitons appear in the spectra of both crystal structures.
These are signatures of the strong \textit{e-h} interaction in organic crystals, also beyond the first few excitations. 
This is also confirmed by the spectra shown in Fig. \ref{fig3}b (upper panel).
By comparing \textit{triplet} (green bars) and $H_0^{BSE}$ excitation energiess (turquoise shaded area), we observe a large difference between their absorption onsets, similarly to the case of the 1T molecule. 
Also LFE play an important role. 
The \textit{singlet} spectrum (blue line) is blue-shifted by over 2 eV compared to the \textit{triplet} onset and the OS of the low-energy peaks is drastically reduced with respect to $H_0^{BSE}$.
Given this complexity, it is not surprising that TDDFT presents serious problems in correctly yielding the spectra.
ALDA slightly underestimates the absorption onset (Table \ref{table1}) and gives only two peaks in the respective energy regime, as seen in Fig. \ref{fig3}b (bottom panel). 
Likewise, two intense peaks appear also in the IPA spectrum (gray shaded area), $\sim$ 1 eV below the ALDA onset (orange line).
In both spectra, the continuum starts at about 8 eV, in a region where BSE features bound excitonic peaks. 
This confirms once again that ALDA reproduces only ``IPA-like" excitations, and thus cannot quantitatively capture the spectral features of molecular crystals.

\section{Summary and Conclusions}
In summary, through a systematic analysis of the optical absorption features in oligothiophenes, we have clarified the role of $e$-$h$ exchange and correlation in describing optical excitations in molecular materials.
For the crystal structures, TDDFT turns out to be an inadequate approach, being unable to reproduce bound excitons and drastically underestimating the continuum onset.
In large oligomers, as shown for the case of 6T, correlation effects turn up in terms of \textit{e-h} interaction, evidenced by mixed excitations. Conversely, in the single thiophene ring, the low-energy excitations are driven by LFE. Neither scenario is captured by TDDFT.
ALDA results are in excellent agreement with BSE merely in the case of 2T, where the first intense peak stems from one vertical transition. 
From this we conclude that ALDA can be trusted for optical excitations in molecular systems barely when correlation effects do not play a predominant role. 
Only high-level methodologies, such as MBPT, which explicitly take into account $e$-$e$ and $e$-$h$ interactions, can reproduce optical spectra of organic materials, from the gas phase to any of their crystalline structure on equal footing.

\section*{Acknowledgment}
Lucia Reining and Kieron Burke are gratefully acknowledged for stimulating discussions.
C.~D.~ thanks Hardy Gross for asking \textit{the} question that inspired this work.
This work was partly funded by the German Research Foundation (DFG), through Collaborative Research Centers SFB-658 and SFB-951.


\end{document}